# Spin-Layer- and Spin-Valley-Locking in CVD-Grown AA'- and AB-Stacked Tungsten-Disulfide Bilayers

Lorenz Maximilian Schneider[1], Jan Kuhnert[1], Simon Schmitt[1], Wolfram Heimbrodt[1], Ulrich Huttner[1], Lars Meckbach[1], Tineke Stroucken[1], Stephan W. Koch[1], Shichen Fu[2], Xiaotian Wang[2], Kyungnam Kang[2], Eui-Hyeok Yang[2], and Arash Rahimi-Iman*[1]

1) Faculty of Physics and Materials Sciences Center, Philipps-Universität Marburg, Marburg 35032, Germany
2) Department of Mechanical Engineering, Stevens Institute of Technology, Hoboken, NJ 07030, United States of America

**Abstract:**

Valley-selective optical selection rules and a spin-valley locking in transition-metal dichalcogenide (TMDC) monolayers are at the heart of "valleytronic physics", which exploits the valley degree of freedom and has been a major research topic in recent years. In contrast, valleytronic properties of TMDC bilayers have not been in the focus so much by now. Here, we report on the valleytronic properties and optical characterization of bilayers of WS₂ as a representative TMDC material. In particular, we study the influence of the relative layer alignment in TMDC homo-bilayer samples on their polarization-dependent optical properties. Therefore, CVD-grown WS₂ bilayer samples have been prepared that favor either the inversion symmetric AA' stacking or AB stacking without inversion symmetry during synthesis. Subsequently, a detailed analysis of reflection contrast and photoluminescence spectra under different polarization conditions has been performed. We observe circular and linear dichroism of the photoluminescence that is more pronounced for the AB stacking configuration. Our experimental findings are supported by theoretical calculations showing that the observed dichroism can be linked to optical selection rules, that maintain the spin-valley locking in the AB-stacked WS₂ bilayer, whereas a spin-layer-locking is present the inversion symmetric AA' bilayer instead. Furthermore, our theoretical calculations predict a small relative shift of the excitonic resonances in both stacking configurations, which is also experimentally observed.

The ability to obtain van-der-Waals (vdW) materials as monolayers has rendered them an emerging novel material class. Transition metal dichalcogenides (TMDCs) as two-dimensional semiconductors have attracted considerable attention, because of their extraordinary strong light-matter interaction and excitonic effects, but also because of their potential application in valleytronic devices.

Common to TMDCs and other layered vdW materials is the honeycomb geometry in real space as well as in reciprocal space with direct band gaps occuring at the corners of the Brillouin zone. Opposite corners are related by the parity or time-reversal symmetry and are referred to as *K* and *K'* valleys. The broken spatial inversion symmetry in a TMDC monolayer allows addressing opposite valleys separately with circular polarized light. This opens up the possibility of using the valley index – or pseudo-spin – for information storage and processing, to name but a few applications.

Many optical investigations on TMDC monolayers have focused on the aspect of valley coherence and found a pronounced optical helicity [1-7] as well as a linear-polarization anisotropy for emitted light after polarized excitation. [8-10] Additionally, control and manipulation of the valley polarisation of TMDC monolayers by means of magnetic [9, 11-15] and electric fields [16, 17] have been demonstrated, and, as has been shown only recently, excitation with strong THz-radiation can be used to control the valley pseudo-spin on a femtosecond time scale. [18, 19]

Meanwhile, recent advances in growth and stacking technology have shifted the focus towards few-layer systems, [2, 20-28] such as systematically arranged hetero- and homo-bilayers. Being able to combine different vdW-Materials in a "lego"-building-block fashion [29] gives numerous possibilities in terms of tailoring the optoelectronic properties of heterostructures. [30]

Within the context of heterostructuring, interlayer excitons with an enhanced radiative lifetime have been predicted theoretically [31, 32] and observed experimentally, not only in hetero-bilayers with type-II band alignment [29, 33-40], but also in homo-bilayers. [31, 41] However, regarding the optical selection rules and hence the addressability of distinct valleys, the overall symmetry and alignment of layers become crucial. While the first studies on bilayers (BL) mainly targeted the naturally-occurring AA'-stacked type of bilayer, only very few studies exist which study the influence of the alignment in artificially stacked homobilayers. [2, 20-23, 42, 43]

Naturally, group VI-TMDCs occur in the inversion symmetric AA' (2H) stacking configuration, or seldom in the AB (3R) stacking configuration without inversion symmetry. The missing inversion symmetry for the AB stacking configuration gives, on the one hand, rise to a non-zero second-order susceptibility $\chi_2$, on the other hand, valley specific selection rules should still persist, rendering this configuration a possible candidate for valleytronic applications.

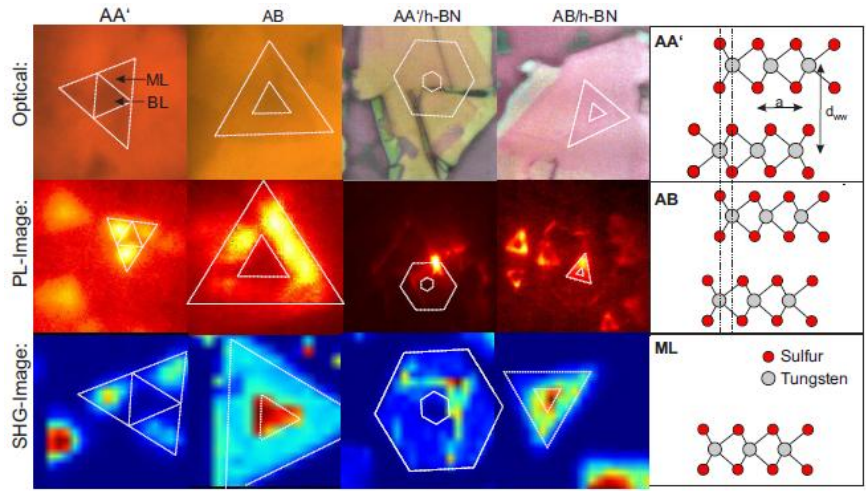

**Figure 1 | Optical characterization overview for differently stacked bilayers:** *Microscopical images of the investigated samples are shown (1st row) together with corresponding false-color PL (2nd row) as well as raster-scan SHG micrographs (3rd row). The white dashed lines are guides to the eyes and are indicating the edges of the WS$_2$ layers in the BL structure. A schematic view of the atomic structure for both stacking configurations is given on the right.*

In this paper, we study the polarization-dependent optical properties of AA' and AB WS$_2$ bilayers. The experiments have been performed on samples grown *via* a low-pressure chemical vapor deposition (LPCVD) process, which enables the controlled growth of bilayers with defined stacking configuration.

In our investigations including SiO$_2$- and hBN-supported (hBN for hexagonal boron nitride) BLs of both types, we experimentally evidence differences in their circular as well as linear dichroism of photoluminescence, which becomes even more pronounced for the hBN-buffered samples. Additionally, we find slight changes in the excitons' energetic position (and the Raman spectra). The observed changes in reflection and emission are in agreement with theoretical calculations based on a combination of density-functional-theory (DFT), gap equations (GE) and Dirac-Bloch equations (DBE) included in this work.

## Samples

AA' and AB BLs have been studied which were grown by low-pressure CVD and afterwards transferred onto a bare $SiO_2$ surface or onto a 2D-buffer structure made of multilayer h-BN on the same substrate material (cf. **Fig. 1** for an overview on the samples). A detailed description of the sample growth and preparation can be found in the Method section. A microscopic picture is shown in the top row of **Fig. 1**. To verify the alignment of the monolayers composing the respective BL stack relative to each other, that is the stacking configuration, photoluminescence (PL) (second row of **Fig. 1**, **Fig. SI.3**), second-harmonic generation (SHG) images (third row of **Fig. 1**) as well as Raman scans (**Fig. SI.2** and **Fig. SI.3**) and TEM diffraction patterns (**Fig. SI.1**) have been acquired. A significant decrease of PL-intensity is found in the BL region as well as a destructive respectively constructive SHG behaviour is observed for AA' and AB, respectively, as expected. The SHG signal confirms the inversion symmetry of the AA' stacked cases and the symmetry breaking in the AB-type samples. The emission peaks as well as the Raman signatures are in agreement with previous studies [44, 45] and confirm the bilayer stacking. Considerations regarding the homogeneity of the optical properties for these samples can be found in the Supporting Information.

## Theory

### *DFT Computations*

We compute the bandstructure and dipole-matrix elements of the different stacking configurations of the WS₂ heterobilayers in an *ab-initio* approach *via* density functional theory (DFT) [46] utilizing the *Vienna ab initio simulation package* (VASP) [47-49] including the spin-orbit interaction. Details and simulation parameters are provided in the Method section.

The resulting DFT band structures for the two different stacking configurations (AA', AB) are shown in **Fig. 2**(e, f) together with the band structure for the WS₂ monolayer (d). The colors represent the spin orientations *up* (red) and *down* (blue) associated with each band.

In case of an isolated monolayer, a direct gap occurs at the $K$- and $K'$-points of the Brillioun zone, and two relevant pairs of bands with opposite spin exist close to the bandgap. As the $K$- and $K'$-points are related by the parity or time-reversal symmetry, the spin projection of the upper and lower spin-split bands are interchanged in the two distinct valleys, and optical transitions are possible between bands of identical spin orientation, as schematically shown in **Fig. 2**(a); similar sketches for both discussed bilayer configurations are shown in (b) and (c) and are discussed below. For the allowed interband transitions in the monolayer, we find the DFT transition energies of 1.627 eV and 2.022 eV, corresponding to the A- and B- exciton series. Due to the lack of inversion symmetry, the $K$-point ($K'$-point) can only be excited by $\sigma^-$ ($\sigma^+$) polarized light. The corresponding dipole moments are summarized in **Table 1**.

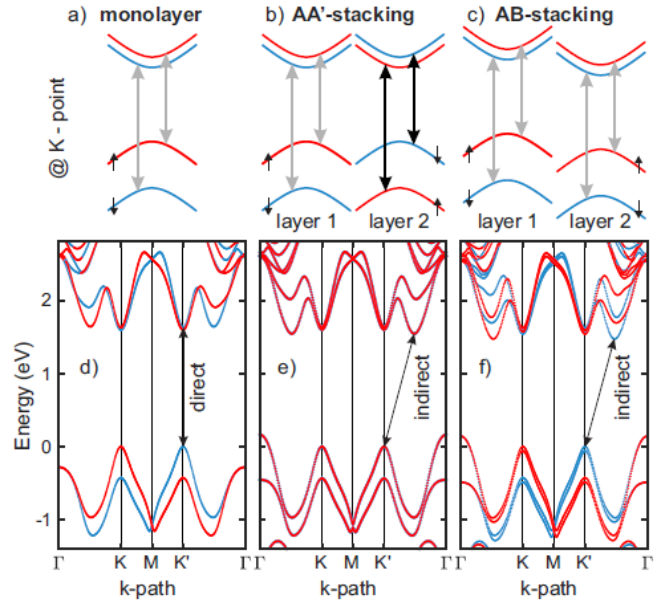

***Figure 2 | Band structure of WS₂ in different stacking configurations:*** *(a) - (c) Schematic representation of the band dispersion at the K point for the ML (a), AA'-BL (b) and AB-BL configuration (c), where the bands are attributed to one of the layers, according to their dominant contributions. The colors represent the spin orientations up (red) and down (blue) associated with each band, while the vertical arrows represent an excitation with σ⁻ (black) and σ⁺ (grey) polarized light. (d) - (f) Full band structure of the ML (d), AA'-BL (e) and AB-BL configuration (f) along the Γ−K−M−K'−Γ path.*

Comparing the band structure of the bilayer configurations with the monolayer band dispersion, we observe strong de*via*tions around the Γ-point and along the Γ−K path of the joint Brillioun zone, indicating an appreciable hybridization and layer mixing in both stacking configurations. In contrast, the shape of the bands around the *K*-points is very similar to the monolayer band dispersion. From this, it becomes evident that the electronic structure of both bilayer configurations around the *K*- and *K'*-points is in very good approximation defined by the superposition of the two MLs' bands, however with very distinct symmetry properties for the two different stacking configurations.

In the AA' stacking configuration, two mirror-identical copies of monolayers are stacked on top of each other, yielding an overall inversion symmetry for the bilayer structure. Consequently, the *K* and *K'*-points of the joint Brillioun zone are equivalent in this stacking and, according to time-reversal symmetry, can be excited with light of either left- or right-handed circular polarization. In reciprocal space, the AA' stacking maps the *K* and *K'*-points of the individual layers onto a single *K*-point of the joint Brillioun zone. Hence, the inversion symmetry leads to the emergence of two pairs of degenerate bands *at each K-point*, which can be distinguished by their respective spin and layer projection as schematically depicted in **Fig. 2**(b).

While optical transitions strictly conserve spin, interlayer interactions also enable transitions between bands dominantly localized within different layers. However, the interlayer interactions are weak as a result of the spin splitting of the bands. The dipole moments associated with these interlayer transitions are about one order of magnitude smaller than the intralayer transitions and, therefore, those transitions will be neglected in the following. Nevertheless, the small interlayer interactions lead to a small difference for the transition energies as compared to those of the monolayer. The DFT transition energies at the *K*-point are summarized in **Table 1** together with the magnitude of the respective dipole-matrix elements.

In an AA or AB stacking configuration, two monolayers with identical orientation are placed on top of each other, where the AB stacking differs from the AA stacking by a relative shift of the monolayer lattices. As for the monolayer, this stacking configuration lacks inversion symmetry and two enantiomers exist. In reciprocal space, stacking two monolayers with identical orientation maps the equivalent $K$ ($K'$)-points of the individual layers onto the corresponding $K$ ($K'$)-point of the joint Brillioun zone. Consequently, the upper and lower spin split bands at a given valley have identical spin and excitation with circular polarized lights leads to a selective valley excitation as for the monolayer case, shown schematically in **Fig. 2**(c). However, the relative shift of the individual layers allows distinguishing the upper and lower layer, such that the bands around the $K$-points can be unambiguously attributed to a specific layer. The distinguishability between the layers lifts the degeneracy between the bands localized within the individual layers and the bands have slightly different energies at the $K$-point. Explicitly, we find the DFT transition energies of 1.630 eV or 1.622 eV (lowest and second lowest band) for the A-exciton transitions and 2.025 or 2.017 eV for the B case.

***Table 1 | DFT transition energies:*** *Transition energies and magnitude of the dipole matrix elements of the different stacking configurations at the K point compared to the monolayer values. Note: The DFT transition energies do not include the band-gap renormalizations due to Coulomb interaction effects.*

| | ML | AA'-BL | AB-BL |
|---|---|---|---|
| $X_A$ energy (eV) | 1.627 | 1.634 | 1.622 / 1.630 |
| $X_B$ energy (eV) | 2.022 | 2.035 | 2.017 / 2.025 |
| $d_A$ (eÅ) | 4.00 | 3.79 | 4.01 / 4.00 |
| $d_B$ (eÅ) | 3.09 | 2.97 | 3.09 / 3.09 |

## Optical Spectra

To compute the optical spectra, we employ our recently developed theoretical framework that combines an electrostatic model for the Coulomb interaction potential in an anisotropic medium, the gap equations (GE) to determine the interacting gap, and the Dirac–Bloch equations (DBE) to compute the linear optical response. [32, 41, 50] For each spin component, we include the two highest valence and two lowest conduction bands near the $K$ points of the joint Brillouin zone, using the DFT transition energies and dipole-matrix elements as input parameters.

According to our DFT analysis, in both stacking configurations, the near $K$-point valence and conduction bands can in good approximation be attributed to a given layer. Hence, the electronic and optical properties around the $K$ points can be treated as those of electronically-independent layers that are coupled *via* the Coulomb interaction and the optical field.

For each layer, we employ the massive-Dirac-fermion (MDF) model's Hamiltonian [51] with a symmetry-induced spin and pseudospin locking of the adjacent layers. Within this effective four-band model, screening of the explicitly treated bands is included dynamically, whereas screening of all other bands and the dielectric environment is obtained as a solution of Poisson's equation for the anisotropic environment, composed of the substrate, capping layers, and non-resonant contributions of the WS$_2$ bilayer.

Since DFT-based band-structure calculations usually underestimate the optical band gap, in a first step, we compute the gap renormalization self-consistently *via* the GE. These GE are a set of coupled integral equations for the renormalized band gap and the Fermi-velocity and can be derived as static solution of the equations of motion in the absence of an external field [50, 52-54]. Subsequently, the linear susceptibility is computed from the DBE and the reflection contrast spectra from Maxwell's equations for the optical field.

**Table 2 | Comparison of theoretical and experimental excitonic resonances:** *Transition energies as computed within our theory framework for the different stacking configurations on a fused silica substrate, where layer 2 is facing towards the substrate. For comparison, the extracted excitonic energies of A and B excitonic series in eV, as obtainable from the reflections contrast measurement, are given. The experimental obtainable values from our measurement that averages over left and right circular light, are written in italic. The peak with lower energy is attributed to the layer with lower energy in the theory model, that means in one case layer 1 and in one case 2.*

| Stacking | layer | gap A (eV) | A-exciton resonances (eV) (Theo.) | (Exp.) | gap B (eV) | B-exciton resonances (eV) (Theo.) | (Exp.) |
|---|---|---|---|---|---|---|---|
| Monolayer | # 1 | 2.467 | 2.123/2.338/2.399 | 2.123/2.230 | 2.940 | 2.554/2.786/2.855 | 2.525 |
| AA'-BL | # 1 | 2.356 | 2.100/2.256/2.300 | 2.103/2.230 | 2.811 | 2.539/2.709/2.757 | 2.507/2.626 |
| | # 2 | 2.328 | 2.089/2.234/2.275 | 2.085/2.198 | 2.807 | 2.533/2.695/2.742 | 2.489/2.623 |
| AB-BL ($\sigma^-$) | # 1 | 2.327 | 2.082/2.223/2.275 | *2.099/2.227* | 2.792 | 2.516/2.682/2.729 | *2.512/2.628* |
| | # 2 | 2.322 | 2.084/2.229/2.270 | *2.082/2.227* | 2.780 | 2.514/2.672/2.717 | *2.512/2.620* |
| AB-BL ($\sigma^+$) | # 1 | 2.337 | 2.091/2.242/2.284 | *2.099/2.227* | 2.801 | 2.524/2.691/2.737 | *2.529/2.628* |
| | # 2 | 2.313 | 2.075/2.220/2.261 | *2.079/2.227* | 2.777 | 2.510/2.669/2.714 | *2.509/2.620* |

The resulting optical spectra depend on the dielectric environment *via* its implicit dependence of the Coulomb matrix elements. In particular, the presence of a substrate leads to a layer-dependent gap renormalization, exciton binding and oscillator strength. For the inversion symmetric AA' stacking, the layer-dependent Coulomb interaction potential leads to a lifting of the degeneracy. The corresponding small splitting of the dipole-allowed exciton transitions can, in principle, be observed under arbitrary excitation conditions. In contrast, the small splitting of the DFT bands associated with different layers in the AB stacking configuration can be either enhanced or reduced by the layer dependent Coulomb renormalizations, leading to a circular dichroism. A detailed list of the computed transition energies on a fused silica substrate is given in **Table 2** and the computed reflection contrast spectra upon excitation with linearly polarized light are shown in **Fig. 3**(b). For the reflection contrast spectra, calculations have been performed using a phenomenological dephasing rate of 30 meV for the A-exciton and 45 meV for the B-exciton. Clearly observable is a red shift of the resonances in the AB stacking configuration as compared to the AA' stacking. As linear excitation conditions average over all dipole-allowed transitions, the calculated splitting for the AB-stacked bilayer can be hardly resolved for the experimental dephasing rate but is barely recognizable as slight asymmetry in the lineshape.

Therefore, we calculate the transition energies for the AB-stacked bilayer for different circular polarizations. The results are summarized in **Table 2** and clearly demonstrate a splitting of the exciton resonance of 16 meV, which is strongest for excitation with $\sigma^+$-polarized light. Moreover, we observe a clear anisotropy. At this point, it should be noted that the calculations have been performed for one of the two possible enantiomeric forms, that in principle may be formed with equal probability and, for the other enantiomer, circular polarizations are reversed.

## Experiment

### Reflection Contrast

In order to compare experimental results with the theoretical predictions, reflection-contrast spectra were measured for several flakes of the same type. Almost no change was observed for different flakes. The derivative for 10 K spectra is shown in **Fig. 3** in comparison with the theoretical predictions. One can clearly observe the 1s and 2s resonance of the A- and B-exciton series. The measurement is in good agreement with the theoretical predictions, as the resonances match almost perfectly in position; especially important, the red shift of the AB bilayer compared to the AA'-stacked layer can be confirmed.

Furthermore, the reflection contrast (a zoom-in to the A-exciton is shown in the inset of **Fig. 3**(a)) and its derivative show a clear asymmetric line-shape for both bilayer types at the main A-exciton resonance as well as a slight asymmetry at the main B-exciton resonance compared to the monolayer. However, the splitting cannot be resolved for the higher-order modes due to limitations set by the linewidth. The energy difference is in agreement with the predicted lifted degeneracy of both layers due to the anisotropic dielectric surrounding, but it appears to be slightly bigger than predicted. The extracted resonance positions are given in **Table 2**.

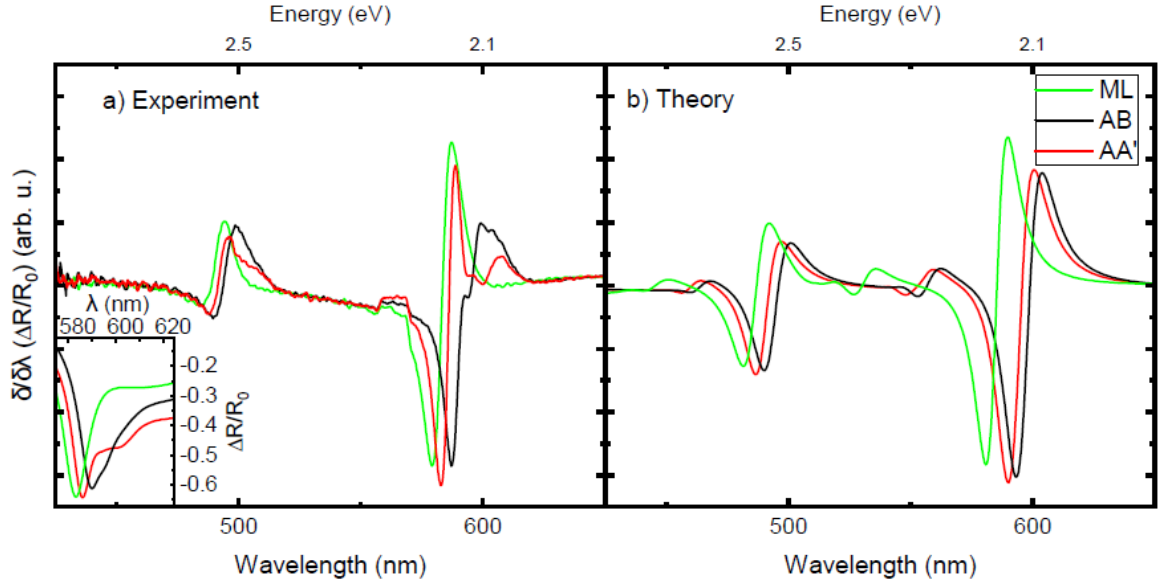

**Figure 3 | Reflection contrast spectra for WS₂ bilayers on SiO₂ in different stacking configurations:** *(a) Experimental derivative of the reflection contrast at 12 K. The inset shows a zoom-in of the reflection contrast at the A-exciton resonance. (b) Calculated derivatives of the reflection contrast spectra for AB stacking (black line) compared to the AA'-stacked bilayer (red line) upon excitation with linear polarized light using a phenomenological dephasing rate of 30 meV for the A-exciton and 45 meV for the B-exciton resonance. The monolayer (green line) is always included for reference.*

## Circular and Linear Dichroism

The co- and contra-polarized PL spectra at 12 K for both bilayer types on bare SiO₂ as well as on an h-BN buffer are shown in **Fig. 4**. The samples were excited quasi-resonantly at 532 nm. Therefore, only the A exciton can be excited. One can clearly observe a circular dichroism (**Fig. 4**(a)) as well as a linear anisotropy (**Fig. 4**(b)). It can be found for excitons and trions, but not for localized-exciton states. However, the amount of anisotropy is different for these configurations. It is bigger for the AB-stacked bilayer compared to the AA' stacked type. The degree of anisotropy is increased on top of the h-BN buffer as expected, due to the screening of the substrate by an atomically flat, uncharged buffer. The extracted degree is calculated as $\rho = (I_{Co}-I_{Contra})/(I_{Co}+I_{Contra})$ and is given in **Table 3** for circular and linear anisotropy. **Fig. 4**(c) and (d) show a helicity-resolved power series for, both, an AA' and AB stacked bilayer under circular excitation. The amount of dichroism (cf. insets of **Fig. 4**(c,d)) is not changing with the excitation density in the used density range.

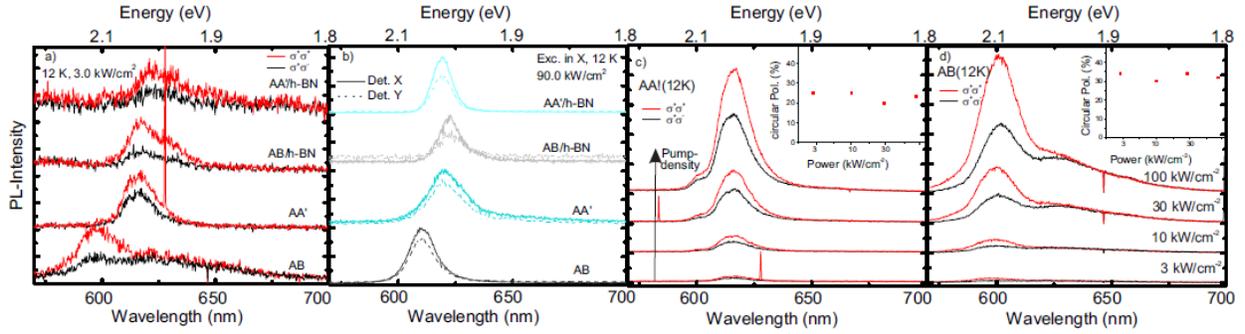

***Figure 4 | Circular and linear dichroism:** (a) Comparison of the PL spectra at 12 K for co- and counter-polarized helicities, regarding excitation and detection polarization. In order to avoid density-related effects and carrier–carrier scattering, the pump density was kept at 3 kW/cm². The AB bilayer shows a stronger anisotropy than its AA' counterpart. (b) Comparison of the linear polarisation anisotropy for the different configurations. The difference in polarization is known to be related to the coherence of the circular dichroism. [55] (c) and (d) show PL helicity for bilayers on SiO₂ as a function of the incident optical power for the AA' and AB case, respectively. The insets show the density-dependent extracted degree of circular polarization at the main feature of the corresponding bilayers.*

***Table 3 | Summary of the magnitude of linear and circular dichroism at 12 K.***

|               | AA' | AB | AA'/hBN | AB/h-BN |
|---------------|-----|-----|---------|---------|
| circ. Pol. (%) | 25  | 34  | 35      | 50      |
| lin. Pol. (%)  | 11  | 12  | 21      | 32      |

## Discussion

As discussed in the theory section, circular dichroism is expected for both bilayer stacking types. However, its origin is different. It is linked to spin-layer locking for the AA' configuration and to spin-valley locking in the AB type. The steady-state degree of circular polarization is a measure how stable the initially excited population in one valley (or layer) is. Complementary, the amount of linear anisotropy in this scenario is linked to the coherence of the initial population distribution. [9] The experimental findings of linear and circular anisotropy therefore confirm the theoretically predicted selection rules.

Particularly, the evidence of inversion symmetry in the AA' case by the absence of SHG while at the same time circular dichroism can be found, makes it clear that the circular anisotropy cannot be explained by a valley polarization. Hereby, we are indirectly evidencing a new type of selection rules as predicted by our calculations. In principle, a layer-resolved photocurrent/photovoltage measurement could in the future directly confirm our findings. The high amount of linear anisotropy, that was found in a steady-state measurement, indicates a coherence time of the population on a similar timescale to the case of a monolayer. [9, 56]

The different strength of anisotropy found for both sample types (AA' and AB on SiO₂ and on h-BN) could reveal a different coherence time of spin-layer- compared to spin-valley-locking for these samples. The scattering processes, that destroy the coherence, differ for both configurations. Therefore, in the AA'-stacked case, a scattering process to the other layer involving a spin flip is needed. In contrast, for the AB bilayer, phonon scattering to the other valley involving a spin flip is necessary in order to result in a counter-polarized emission. The higher degree of dichroism in the AB stack suggests that the relevant scattering processes are more likely for the AA' than for the AB configuration, at 12 K.

Nevertheless, as both types show a locking of the spin with a non-negligible coherence time, both stacking types can be useful in certain situations. The AA'-stacked bilayer, on the one hand, can be used in a "layertronics" device, [57, 58] a concept recently proposed for a bilayer of graphene, that can be adressed with circular light. In contrast to graphene, the light–matter interaction in WS$_2$ is much stronger, making the AA'-stacked bilayer a potential building block for light sensing in a layertronic design. The AB-stacked bilayer, on the other hand, can be used in valleytronics like its monolayer pendant. However, as an indirect semiconductor, this bilayer type can be used as additional building block if radiative losses are a concern in valleytronic design or a valleytronic material at the indirect gap is needed. Furthermore, if only one enantiomer is used, a controlled excitation of only one of the layers (and one valley) is in principle possible, opening even more possible concepts that may pave the way for combining valley- and layertronics. While in general the symmetry considerations are true for other TMDCs, they differ strongly in their interlayer interaction dipole moments. Thus, it can be expected that not all homobilayers of TMDC can be used for that purpose.

## Summary

In conclusion, we have studied the two stable stacking types of WS$_2$ bilayers, namely AA' and AB, that differ strongly in symmetry. These changes in symmetry have certain implications on the energetics and their optical selection rules. Our theoretical calculations have shown that AA' bilayers exhibit indeed a spin–layer locking rather than a spin–valley locking that is present in the AB configuration. Both of them lead to a circular dichroism that could be obtained experimentally, while having measured an anisotropy and polarisation coherence bigger for the AB stack than for the AA'-bilayer type. Yielding circular dichroism of PL, while lacking SHG signal from the AA' stack, indicates that the origin can indeed not be explained by valley polarization, but has to be considered as indirect evidence of spin–layer locking. Furthermore, a lifting of the degeneracy of both layers predicted by our calculations could be seen experimentally, making it possible to identify or address a single layer of the bilayer stack in both cases, if necessary. The observed spin locking and the possibility to only adress one layer opens up new possibilities in layer- respectively valleytronics or even make it possible to combine them.

## Methods

**Theoretical Details:** The band structure and dipole-matrix elements of the WS$_2$ homobilayers are computed in an *ab-initio* approach *via* density functional theory (DFT) [46] utilizing the *Vienna ab initio simulation package* (VASP) [47-49] including the spin-orbit interaction. [59] VASP is a projector-augmented plane-wave code, such that only the relevant valence electrons of an atom have to be fully treated, while the lower lying electrons are included in the ionic core potential. [60, 61] The exchange-correlation potential is treated *via* the generalized gradient approach as parametrized in the Perdew-Burke-Ernzerhof (PBE) functional. [62] Additionally, the van-der-Waals interaction between the individual layers is included *via* Grimme's dispersion correction method (PBE-D3), [63, 64] which accurately reproduces the bulk lattice constants of WS$_2$ and other TMDCs. The reciprocal space is sampled by a 12x12x3 Monkhost-Pack *K*-mesh, [65] centered around the Γ point, where the homobilayers are aligned in the *xy*-plane. The plane-wave basis-set includes wavevectors up to the cut-off energy of 750 eV for structural relaxations and 260 eV for computations of the electronic properties.

Considering freestanding bilayers, the unit cells of both stacking configurations under investigation contain six atoms each. To prevent artificial interactions between the individual periodic copies of the unit cell in *z*-direction, a vacuum region of 20 Å is added in between the bilayers. All structures are fully relaxed until the resulting inter-atomic forces are smaller than 0.0025 eV/Å. To obtain accurate dipole-matrix elements, especially in the case of the AA' stacking configuration, where degenerate bands exist, we perform the electronic minimization until the energy convergence criterion of at least $10^{-7}$ eV is reached, *via* an exact diagonalization of the one-electron Hamiltonian. This approach also includes the relevant, empty conduction bands, which do not contribute to the total energy and, therefore, would otherwise not necessarily be converged. The dipole-matrix elements are then computed *via* the linear optics package [66] available in VASP.

**Experimental Details:**

**Sample Growth and Preparation:**

WS₂ homobilayers were synthesized *via* the two-step low-pressure chemical vapor deposition (LPCVD) process. In general, WS₂ monolayers obtained from the first growth step will provide seeding sites for another monolayer of WS₂ to grow on top during the second run, resulting in a WS₂ homobilayer. First, a source substrate was prepared by depositing 5 nm tungsten trioxide on a SiO₂/Si or sapphire substrate using physical vapor deposition (PVD). The source substrate was placed atop another SiO₂/Si or sapphire substrate (i.e., growth substrate) face-to-face. During the first growth, WS₂ monolayers were grown on the growth substrate at 950 °C. In the second growth, a new source substrate was placed face-to-face onto the growth substrate (with WS₂ monolayers formed during the first step growth).

The WS₂ monolayers grown during the first step provided seeding sites for WS₂ during the second growth at 850 °C, resulting in growing WS₂ homobilayers with an AA' or AB stacking mode. These two different stacking modes were determined by adjusting the growth parameters of WS₂ monolayers obtained during the first step growth. The amount of sulfur can be adjusted during the growth, which favors either a sulfur rich environment or a tungsten trioxides rich environment. The sulfur rich environment was found to give sulfur-terminated WS₂ monolayers, [67] which triggered the AB stacking of the homobilayers, while the tungsten trioxides rich environment was found to give tungsten-terminated WS₂ monolayers, [67] which triggered the AA' stacking of the homobilayers.

**Sample Transfer**: AA' and AB WS₂ BLs grown on SiO₂ substrate was covered with PMMA (950 A4) by a dropper. The PMMA-covered sample was left in an ambient condition to dry PMMA for 60 min, and was subsequently floated on 30% KOH (aq) for 3-5 min, which separated the PMMA/WS₂ layer from the growth substrate. Next, the PMMA/WS₂ layer was cleaned in DI water for 5 times to remove KOH residue. Cleaned PMMA/WS₂ layer was attached to the target substrate (i.e. SiO₂ or hBN), dried in ambient conditions for 1 hour, and baked at 90 °C for 1.5 min to enhance the bonding between WS₂ and substrate. PMMA was removed using 50 °C warm acetone for 30 min, followed by an IPA rinsing, DI water rinsing and hotplate baking at 130 °C for 5 min.

**PL-Measurements:** The samples were evacuated in an optical cryostat at high-vacuum condition in an conventional confocal μ-PL setup with a 60x microscope objective (NA 0.7). The confocal spatial selection enables transmission of signal from a sample spot of roughly 1 μm. For reflection-contrast measurements, the reflection of a tungsten lamp was measured with a spectrometer and a nitrogen-cooled camera. For CW excitation and for the SHG measurement, a frequency-stabilized Nd:YAG laser at 532 nm and a pulsed titan-sapphire laser (100 fs) at 900 nm was used, respectively. For polarization-resolved measurements, superachromatic waveplates and polarizers have been used. For hyperspectral PL-scans, the sample was scanned using a piezo stage. For PL images, the laser radiation was filtered by a holographic notch filter and the spatially-resolved signal was recorded using a scientific ICCD.

**Acknowledgement**
This work was supported by the Deutsche Forschungsgemeinschaft (DFG) *via* the Collaborative Research Center 1083 (SFB 1083) and through grant numbers KI 917/3-1 and RA 2841/5-1. Computing resources from the HRZ Marburg are acknowledged.

**Authors' contributions**
E.H.Y. and A.R.-I. initiated the study on CVD-grown bilayers with different stacking symmetry and guided the joint work. The samples were grown and characterized by S.F., X.W. and K.K. with the support of E.H.Y. Microscopically-resolved optical spectroscopy and imaging SHG was performed by L.M.S., J.K., S.S. with the support of W.H. and A.R.-I. Theoretical modelling work was contributed by U.H., L.M., T.S. and S.W.K. The results were discussed with the help of all coauthors. The manuscript was written by L.M.S. and A.R.-I. with strong input from U.H., T.S., K.K. and E.H.Y. All coauthors supported manuscript preparation.

**Corresponding author**
a.r-i@physik.uni-marburg.de

**Associated Content**
The authors declare no conflict of interest

**Supporting Information Available**
Supporting Information Available: Selected area electron diffraction, Linescans of PL and Raman signatures at room temperature that verify the alignment can be found in the Supporting Information.